\documentclass[prb,twocolumn,showpacs,superscriptaddress,amsmath,amssymb]{revtex4}
\usepackage{graphicx}

\usepackage{units}
\usepackage{color}

\usepackage{xcolor}

\def\be{\begin{equation}}
\def\ee{\end{equation}}
\def\ba{\begin{eqnarray}}
\def\ea{\end{eqnarray}}

\usepackage{bbm}

\usepackage{changes}

\begin{document}
	
	\title{Tracing the $s_\pm$-symmetry in iron pnictides by controlled disorder}
	\author{M. B. Schilling}
	\affiliation{1.~Physikalisches Institut, Universit\"at Stuttgart, Pfaffenwaldring 57, 70569 Stuttgart, Germany}
	\author{A. Baumgartner}
	\affiliation{1.~Physikalisches Institut, Universit\"at Stuttgart, Pfaffenwaldring 57, 70569 Stuttgart, Germany}
	\author{B. Gorshunov}
	\affiliation{1.~Physikalisches Institut, Universit\"at Stuttgart, Pfaffenwaldring 57, 70569 Stuttgart, Germany}
	\affiliation{A. M. Prokhorov General Physics Institute, Russian Academy of Sciences,
		%Vavilov str.\ 38,
		119991 Moscow, Russia}
	\affiliation{Moscow Institute of Physics and Technology (State University), 141700, Dolgoprudny, Moscow Region, Russia}
	\author{E. S. Zhukova}
	\affiliation{1.~Physikalisches Institut, Universit\"at Stuttgart, Pfaffenwaldring 57, 70569 Stuttgart, Germany}
	\affiliation{A. M. Prokhorov General Physics Institute, Russian Academy of Sciences,
		%Vavilov str.\ 38,
		119991 Moscow, Russia}
	\affiliation{Moscow Institute of Physics and Technology (State University), 141700, Dolgoprudny, Moscow Region, Russia}
	\author{V. A. Dravin}
	\author{K. V. Mitsen}
	\affiliation{P. N. Lebedev Physical Institute, Moscow, 119991 Russia}
	\author{D. V. Efremov}
	\affiliation{Institute for Theoretical Solid State Physics, IFW Dresden, Germany}
	\author{O. V. Dolgov}
	\affiliation{Max-Planck Institut f\"ur Festk\"orperforschung, 70569 Stuttgart, Germany}
	\affiliation{P. N. Lebedev Physical Institute, Moscow, 119991 Russia}
	\author{K. Iida}
	\affiliation{Institute for Metallic Materials, IFW Dresden, Germany}
	\affiliation{Department of Crystalline Materials Science, Graduate School of Engineering, Nagoya University, Nagoya 464-8603, Japan}
	\author{M. Dressel}
	\affiliation{1.~Physikalisches Institut, Universit\"at Stuttgart, Pfaffenwaldring 57, 70569 Stuttgart, Germany}
	\author{S. Zapf}
	\affiliation{1.~Physikalisches Institut, Universit\"at Stuttgart, Pfaffenwaldring 57, 70569 Stuttgart, Germany}
	\date{\today}

\begin{abstract}
Determining  the symmetry of the superconducting order parameter is the most important, but also the most complicated step in elucidating the mechanism of superconductivity.  Here we present an experimental approach to investigate the order parameter symmetry of unconventional multiband superconductors, which is based on a disorder-induced change from sign-reversed ($s_{\pm}$) to sign-preserved ($s_{++}$) symmetry. Therefore, we investigated a Ba(Fe$_{0.9}$Co$_{0.1}$)$_2$As$_2$ thin film by THz spectroscopy and step-wise proton irradiation. In our experiments, the low-energy superconducting gap first vanishes, but recovers at higher irradiation doses. At the same time, the decrease of the superfluid density with disorder comes to a halt. Thus, we confirm with the here-presented method that the superconducting order parameter in the pristine sample possesses $s_{\pm}$-symmetry.

\end{abstract}
\maketitle

\section{Introduction}
Despite tremendous work by theoreticians and experimentalists during the last years, a universal sign change of the superconducting order parameter between different sheets of the Fermi surface in Fe-based superconductors (FeSC) is still under debate~\cite{Hirschfeld16,rev,rev5,rev2,rev3}. In analogy to cuprates, the emergence of a sharp peak in the dynamic spin susceptibility $\chi''(\mathbf{Q},\omega)$  below the superconducting transition temperature $T_{\text{c}}$ is often considered as strong evidence of such a sign change.
Indeed, this magnetic spin resonance was observed in most FeSC, with very few exceptions like LiFeAs~\cite{Dai15,Inosov16}. However, the peaks found in the experiments are broader than expected. As it was shown that the magnetic spectral weight redistribution due to the opening of a spin gap in $s_{++}$ superconductors  may also cause such a peak below $T_{\text{c}}$, doubt was casted on the unambiguity of the $s_{\pm}$ gap symmetry~\cite{Kontani10,Zhang13}. Hence, its confirmation by an independent experimental method is urgently needed.

Another technique to yield  information about the relative sign of the order parameter  on different sheets of the Fermi surface is scanning tunneling microscopy (STM).
While STM in principle is phase insensitive, the quasi-particle scattering on an impurity strongly depends on the phase of the underlying order parameter \cite{Coleman11,Hirschfeld15, Hanaguri10}. However, these quasi-particle interference (QPI) experiments require a very clean surface. Therefore, they can be done only for a few compounds of FeSC, such as  LiFeAs and FeSe \cite{Hirschfeld16}.

Moreover, doping-induced disorder was discussed as a tuning parameter through the phase diagram of FeSC~\cite{Sawatzky}. In this regard, many experimental observations in doped systems were compatible with an $s_\pm$-superconductor under the influence of disorder~\cite{Noat10,Dai13}. However, controlling the impurity level by step-wise irradiation allows much more direct conclusions. Unfortunately, such experiments concentrated up to now mainly on the suppression of $T_{\text{c}}$~\cite{Nakajima10,Tarantini10,Beek13}, where the inter- and intra-band scattering hampers any definite conclusion~\cite{Efremov11,Wang13}. Similarly, the effect of controlled disorder on the temperature (or energy) dependence of a certain measurement parameter, such as the superconducting penetration depth~\cite{Mizukami14}, may lead to contrary results for differently applied fitting ranges.

In the following, we present an alternative experimental method to investigate the order parameter of multiband superconductors with $s_\pm$ order parameter. Our conclusions are based on the disorder-induced closing- and re-opening of their superconducting gap, which can be directly observed with several experimental techniques, such as optical spectroscopy, STM or angle resolved photoemission spectroscopy (ARPES). From these, THz spectroscopy was our method of choice, as it allows a macroscopic investigation of the bulk properties. Similar to QPI, we use a phase insensitive method in combination with a phase sensitive physical phenomenon to yield information about the sign of the order parameter.

\section{Theoretical background}
Our method is based on a characteristic development of the superconducting gaps in multiband superconductors under the effect of disorder~\cite{Efremov11,Efremov13,Golubov1995,Golubov1997}. It was shown that with increasing disorder, the gap functions on different sheets of the Fermi surface tend to the same value (see Ref.~\cite{Hirschfeld16,Efremov11,Efremov13} and references therein). Since the gap functions have the same sign in $s_{++}$ superconductors, the only possible evolution is a continuous and monotonic approach of all gaps towards the same value. In $s_\pm$ superconductors, the situation is more complicated and admits two scenarios depending on the electron-electron coupling constants~\cite{Efremov11,Efremov13}: in the first scenario, both gaps simultaneously evolve to zero (full suppression of the superconducting phase according to the Abrikosov-Gorkov law); in the second scenario, both gaps tend to the same finite value, inevitably involving the closing and re-opening (with reversed sign) of the smallest gap, thus showing a transition from $s_\pm$ to $s_{++}$. We note that this transition is induced by disorder and not by a change of the interactions responsible for Cooper pairing. The non-monotonic change of the small gap appears only if the pristine system possesses the $s_\pm$-symmetry and can be detected even with a phase insensitive technique. Hence, one does not need any detailed quantitative comparison with theoretical calculations for inevitably proving the $s_\pm$-symmetry with our method.

\section{Experimental background}
In our experiment, we extract the evolution of the small gap $\Delta$ from the coherence peak that appears for multiband systems in $\sigma_1(T, \omega \rightarrow 0)$ at the temperature $T_{\text{max}}$ with  $k_{\text{B}} T_{\text{max}} \sim \Delta$ (see Ref.~\cite{Jin2003} and the Supplemental Material). This peak resembles the Hebel-Slichter peak in nuclear magnetic resonance (NMR) studies~\cite{DresselGruener}, and appears when the parts of the Fermi surface that are coupled by the experimental probe have gaps of the same sign and similar magnitude; thus, it can be observed in the case of Ba(Fe$_{0.9}$Co$_{0.1}$)$_2$As$_2$ by THz spectroscopy~\cite{Aguilar10, Fischer10}. Moreover, it gets suppressed the further a superconductor is in the clean limit~\cite{Nicol}.

\section{Experimental details}
The high-quality Ba\-(Fe$_{0.9}$Co$_{0.1}$)$_2$\-As$_2$ epitaxial film of a thickness of approximately $\unit[50]{nm}$  was grown via pulsed-laser deposition on a CaF$_2$ $(001)$ substrate (thickness $d=\unit[1]{mm}$) and exhibits a sharp superconducting transition with $T_\text{c} = \unit[26]{K}$; details of the growth procedure are described in Ref.~\onlinecite{Iida09}. In order to create non-magnetic impurities~\cite{Nakajima10}, the sample was step-wise irradiated with protons ($D = 2 \times 10^{14}$ protons per cm$^{2}$ for each step) at the DanFysik-350 at the P.N.~Lebedev Physical Institute Moscow, using an energy of $\unit[200]{keV}$. This energy is high enough to guarantee that particles are not implanted, but pass the film and substrate and create point-like defects. By scanning the sample uniformly with a particle-beam of $\unit[2]{mm}$ spot size, the induced defects are distributed homogeneously (better than $\pm1\%$). In total, the film was successively irradiated nine times.

\begin{figure}
	\centering
	\includegraphics[width=0.95\columnwidth]{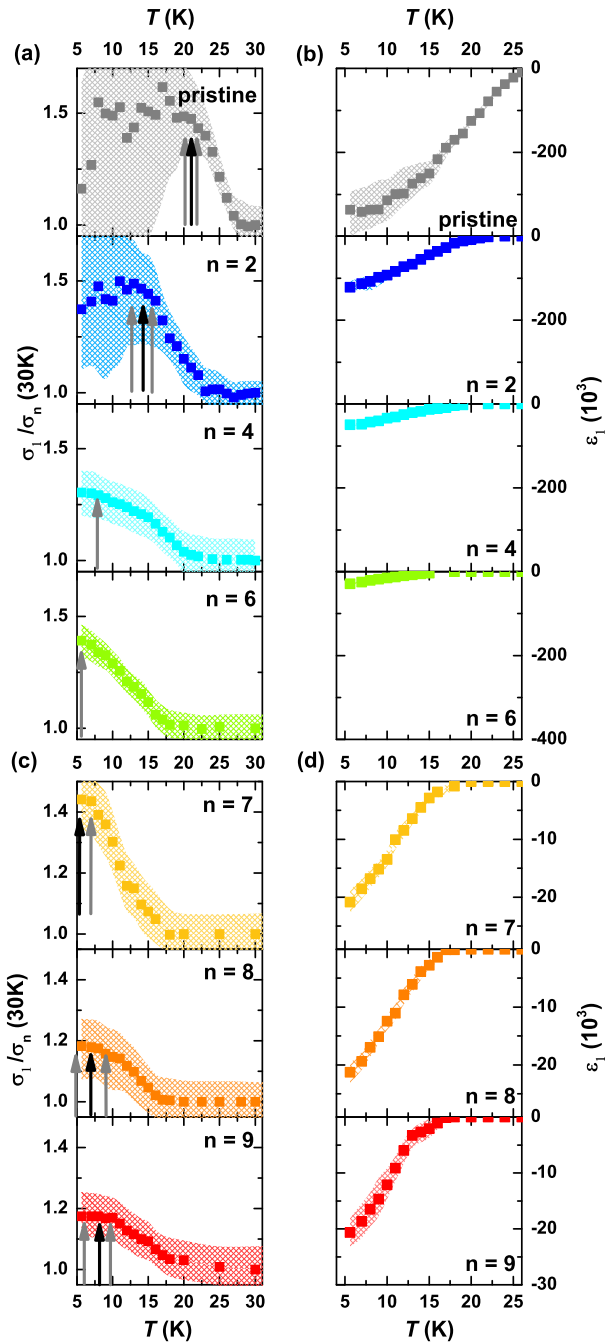}
	\caption{Temperature-dependent optical conductivity $\sigma_1(T)$ normalized to $\unit[30]{K}$ (left) and permittivity $\epsilon_1(T)$ (right) of a Ba(Fe$_{0.9}$Co$_{0.1}$)$_2$As$_2$ thin film at $\nu = \unit[5]{cm^{-1}}$ for different amounts of proton irradiation (dose $D = \unit[\text{n} \cdot 2\times 10^{14}]{cm^{-2}}$). (a)~In the first irradiation cycles, the coherence peak that represents the smallest superconducting gap is shifted to lower temperatures, as indicated by the black arrows (gray arrows represent the uncertainty).
		(b)~Meanwhile, the absolute value of $\epsilon_1$ as well as the temperature of the onset are reduced.
		(c)~After a dose of $\unit[12 \times 10^{14}]{cm^{-2}}$, ($\text{n}=6$), the maximum of the peak is shifted back to higher temperatures.
		(d)~At the same proton dose a change appears in $\epsilon_1(T)$, since the absolute value is not affected anymore by further proton irradiation.
		The shaded area represents the uncertainty in the absolute value.}
	\label{fig12}
\end{figure}
\begin{figure*} [t]
\begin{minipage}[c] {0.7\textwidth}
	\centering
	\includegraphics[width=\textwidth]{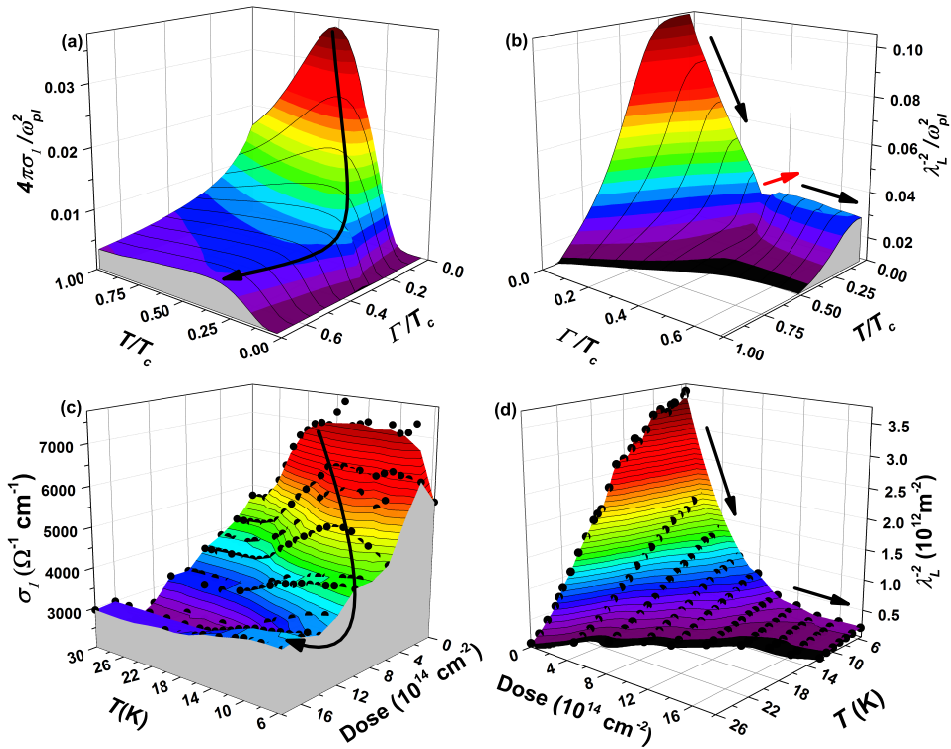}
\end{minipage} \hfill
\begin{minipage}[c] {0.27\textwidth}
	\caption{
Top: Theoretical predictions for (a)~$\sigma_1$ and (b)~$\lambda^{-2}_\text{L}\propto n_s$ as functions of  temperature $T$ and interband impurity scattering amplitude $\Gamma$ in a two-band system with the following parameters: electron-electron coupling constants $\lambda_{11}$ $=3$, $\lambda_{12}$ $=-0.2$, $\lambda_{21}$ $=-0.1$ and $\lambda_{22}$ $=0.5$; scattering rates within in the Born approximation $\Gamma_{11}$ $=10\Gamma$, $\Gamma_{12}/2$ $= \Gamma_{21}$ $= \Gamma_{22}$ $= \Gamma$; the ratio between the plasma frequencies is $\omega_\text{pl,2}^2 / \omega_\text{pl,1}^2 = 2$. These parameters are similar to other calculations for iron pnictides~\cite{Popovich10,Charnukha11}. The transition of the superconducting order parameter from $s_\pm$ to $s_{++}$ can be seen by both, (panel a) the back-shift of the coherence peak to higher temperatures and (panel b) the re-increase of $\lambda^{-2}_\text{L}$ (red arrow) for increasing scattering rate.
Bottom:
Experimentally determined (c)~$\sigma_1$ and (d)~$\lambda^{-2}_\text{L}$ as functions of $T$ and irradiation dose. While the back-shift of the coherence peak can be seen as predicted by theory, the re-increase of $\lambda^{-2}_\text{L}$ at the critical scattering rate $\Gamma_\text{c}$ is not resolved in our results.
	}
	\label{fig34}
\end{minipage}
\end{figure*}

After each step of irradiation, we studied the optical properties of the film by coherent-source THz spectroscopy. Utilizing a Mach-Zehnder interferometer, the transmission coefficient and the phase shift of the electromagnetic radiation were measured at different temperatures down to $T = \unit[5]{K}$ over a frequency range from $\unit[3.5 \text{ to } 36]{cm^{-1}}$ ($\unit[0.41-4.46]{meV}$)
\cite{Gorshunov05}. This allows us to directly determine the complex optical conductivity $\hat{\sigma} = \sigma_1 + \rm{i} \sigma_2 $ of the film without any use of the Kramers-Kronig relation~\cite{Pracht13}.

\section{Experimental results} Discussing our experimental results, we concentrate on the lowest energy ($\nu = \frac{\omega}{2\pi c}=\unit[5.0]{cm^{-1}}$), providing information on the penetration depth $\lambda_\text{L}$, the superfluid density $n_s$, as well as the lowest superconducting energy gap $\Delta$ of the investigated film. More details on the optical spectra can be found in the Supplemental Material \cite{SupplementalMaterials}.
While the former quantities are extracted directly from $\sigma_2(\omega \to 0)$ [ $\epsilon_1(T) =\epsilon_\infty-\frac{4\pi\sigma_2}{\omega}$, with $\epsilon_\infty$ being the optical contribution from the interband transitions], $\Delta$ is gained from the coherence peak that appears in $\sigma_1(\omega \to 0, T)$. Thus, tracing the coherence peak enables us to detect the evolution of this energy gap when disorder increases.

\begin{figure}[t]
	\centering
	\includegraphics[width=1\columnwidth]{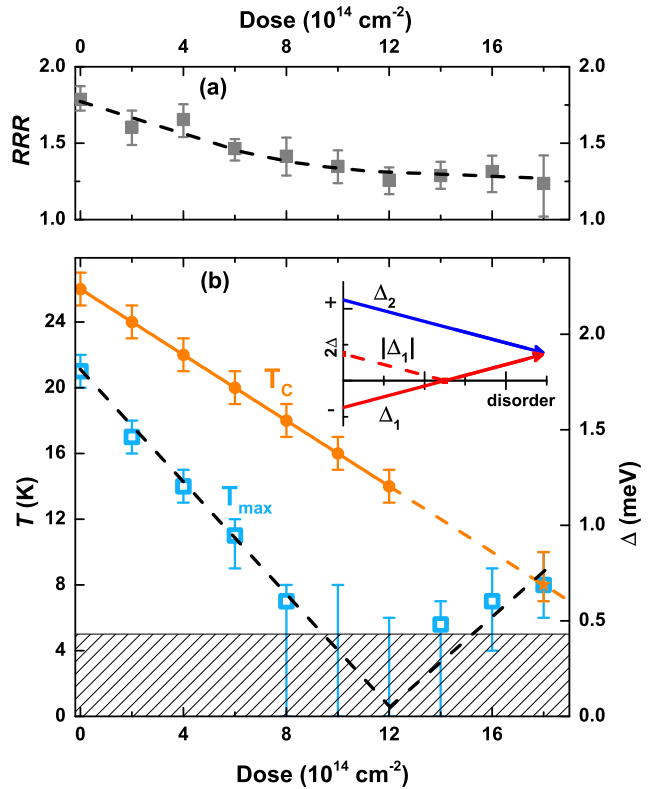}
	\caption{(a)~Residual resistivity ratio of the paramagnetic state determined by the optical conductivity. The reduction of $RRR$ with increasing irradiation proves the accumulating disorder.
(b)~Irradiation dependence of the critical temperature (orange closed circles) and the maximum of the coherence peak (blue open squares) obtained from the optical measurements at $\nu = \unit[5]{cm^{-1}}$. For doses between $\unit[8 \text{ to } 12]{\times 10^{14} cm^{-2}}$, the maximum of the peak drops below the accessible temperature range (shaded area) and we only indicate an upper bound. Starting with a dose of $D = \unit[14 \times 10^{14}]{\,protons\ per\ cm^{2}}$, the maximum of the coherence peak is shifted back into our measured temperature range. The inset sketches the expected behavior of two energy gaps upon increasing disorder. Since they tend to the same value, the smaller gap has changed its sign.}
	\label{fig5}
\end{figure}

The left panel of Fig.~\ref{fig12} shows the development of the coherence peak in the normalized conductivity $\sigma_1(T)$ as a function of irradiation. Note that while the uncertainty in the absolute value (shaded area) of THz frequency-domain spectroscopy measurements is typically rather large in the superconducting state due to the huge inductive response of the Cooper pairs~\cite{THz}, the relative error is significantly smaller. As expected, increasing disorder via proton irradiation decreases not only $T_\text{c}$ (see Fig.~\ref{fig5}), but also reduces the temperature $T_\text{max}$ of the coherence peak maximum (black arrows in Fig.~\ref{fig12}a). After a dose of $D = \unit[8 \times 10^{14}]{cm^{-2}}$, ($\text{n}=4$; $\text{n}$ indicates the number of irradiation), $T_\text{max}$ is shifted to temperatures below the measured temperature window, indicating that the energy gap continues to close.
Here we emphasize that $T_\text{max}$ is suppressed faster than $T_\text{c}$ and hence the BCS relation between the gap value and $T_\text{c}$ is violated (see also Fig.~\ref{fig5}). This is an important prerequisite for the disorder-driven transition from $s_\pm$- to $s_{++}$-symmetry.

Strikingly, at an irradiation dose of $14\times 10^{14}$ $\,\text{protons per cm$^{2}$}$ ($\text{n}=7$), the maximum of the coherence peak moves back into the measured temperature range (see Fig.~\ref{fig12}c), representing the re-opening of the energy gap; further irradiation shifts $T_\text{max}$ to even higher temperatures.

Further evidence for such a transition can be gained from the temperature-dependent permittivity $\epsilon_1(T) \propto \sigma_2(T)$, which is depicted in the right panel of Fig.~\ref{fig12} for different irradiation steps. For the first measurement cycles, the absolute value of the permittivity at lowest temperatures, as well as the temperature where it starts to drop, decrease with increasing disorder (Fig.~\ref{fig12}b), which is consistent with the suppression of the superconducting phase. However, this trend stops after a dose of $\unit[14 \times 10^{14}]{\,protons\ per\ cm^{2}}$ ($\text{n}=7$), and $\epsilon_1(T)$ tends to saturate (see Fig.~\ref{fig12}d). Such a saturation of $\epsilon_1(T)$ corresponds to a saturation of the superfluid density $n_s \propto \lambda^{-2}_\text{L} = 4\pi^2 (\epsilon_\infty-\epsilon_1) \omega^2$, which is depicted in Fig.~\ref{fig34}d. Remarkably, this happens at the same irradiation dose as the back-shift of the coherence peak.

In Fig.~\ref{fig34} we show a qualitative comparison between the theoretical predictions for a two-band model~\cite{Efremov11,Efremov13} and our experimental results for $\sigma_1(T)$ and $\lambda^{-2}_\text{L}(T)$ as a function of the impurity scattering rate $\Gamma$. At $\Gamma = 0$ the theoretical value of the small gap is $\Delta = 0.5 T_{\text{c}}$, which is in good agreement with the experimental observations.  With increasing disorder, the position of the coherence peak, representing the smallest gap, shifts to lower temperatures in both, theoretical calculations and experimental results; this corresponds to the suppression of the small gap $\Delta$.  In the meantime, $\lambda^{-2}_\text{L}$ is decreased due to the suppression of the superconducting phase. 
At a certain critical scattering rate ($\Gamma_\text{c} \approx 0.35T_{\text{c}}$), the trend stops and the peak maximum is shifted back to higher temperatures; this corresponds to the re-opening of the smallest gap with different sign.

While the experimental $\sigma_1(T)$ directly follows the theoretical predictions, there is some difference for $\lambda^{-2}_\text{L}$. Experimentally, we do not resolve the re-increase  of the superfluid density in the vicinity of $\Gamma_\text{c}$. This deviation can be attributed to the experimental resolution, or to the fact that we measure at finite frequencies which reduces the strength of the increase compared to theory. Nevertheless, the leveling off at highest disorder is directly observable.

Figure~\ref{fig5} summarizes our main observations on the basis of the parameters: $T_\text{max} (\propto \Delta)$, $T_\text{c}$ and the residual resistivity ratio ($RRR$) determined from the optical studies~\cite{TcRRR}.
The decrease of $RRR$ and the linear suppression of $T_\text{c}$ indicate an enhancement of the scattering rate that is proportional to the irradiation dose.
As can be clearly seen,  $T_\text{max}$ is suppressed faster than $T_\text{c}$, and hence the BCS relation between gap value and critical temperature is violated, which enables us to observe the transition of the order parameter. For an irradiation dose of $\unit[14 \times 10^{14}]{\,protons\ per\ cm^{2}}$, we clearly see that $T_\text{max}$ shifts back into the measured temperature range.
Further disorder increases $T_\text{max}$ even more, while the reduction of $T_\text{c}$ still goes on, probably with a slightly smaller slope which can be seen from the onset of $\lambda^{-2}_\text{L} (T)$. This represents the re-opening of the smallest gap.
\footnote{While the smaller gap which changes its sign is discussed here, information about the larger energy gap(s) can be found in the Supplemental Material.}

\section{Conclusions.}
In summary, we have presented an alternative experimental approach to investigate the order parameter of unconventional multiband superconductors. Our experimental results show a clear anomaly in the optical properties of a superconducting Ba(Fe$_{0.9}$Co$_{0.1}$)$_2$As$_2$ thin film, when we increase the disorder via proton irradiation: the low-energy superconducting gap gets first suppressed, but recovers at higher irradiation doses ($ D > \unit[14 \times 10^{14}]{\,protons\ per\ cm^{2}}$). At the same time, the decrease of the superfluid density with disorder comes to a halt. Our experimental findings are in excellent agreement with our theoretical predictions~\cite{Efremov11,Efremov13}, which propose a disorder-induced sign-change of the superconducting order parameter from $s_{\pm}$- to $s_{++}$ -symmetry. Thus, our alternative experimental approach confirms that the superconducting order parameter in the pristine Ba(Fe$_{0.9}$Co$_{0.1}$)$_2$As$_2$ sample possesses the $s_{\pm}$-symmetry.\\
\newline
We would like to thank B. Keimer, A. Boris, S.-L. Drechsler, A. Golubov, V. Grinenko, M. Korshunov, A. Kreisel, P. J. Hirschfeld, and I. I. Mazin for useful discussions.

The project was supported by the Deutsche Forschungsgemeinschaft (DFG)
as part of SPP 1458 and
by the Russian Ministry of Education and Science (Program 5 top 100).


\begin{thebibliography}{99}
	
	\bibitem{Hirschfeld16} P. J. Hirschfeld, C. R.Physique {\bf 17},  197	 (2016).
	
	\bibitem{rev2}
	D. C. Johnston, Adv. Phys. {\bf 59},  1063 (2010).
	
	\bibitem{rev3}
	G. R. Stewart, Rev. Mod. Phys. {\bf 83}, 1589 (2011).

	\bibitem{rev} P. J. Hirschfeld, M. M. Korshunov, and I. I. Mazin, Rep. Prog. Phys. \textbf{74}, 124508 (2011).
	
	\bibitem{rev5} H. Hosono and K. Kuroki, Physica C {\bf 514}, 399 (2015).

	\bibitem{Inosov16} D. S. Inosov, C. R. Physique {\bf 17}, 60(2016).
	
	\bibitem{Dai15} P. Dai, Rev. Mod. Phys. {\bf 87}, 855 (2015).
	
	\bibitem{Kontani10} H. Kontani and S. Onari, Phys. Rev. Lett. \textbf{104}, 157001 (2010).
	
	\bibitem{Zhang13} C. Zhang, H.-F. Li, Y. Song, Y. Su, G. Tan, T. Netherton, C. Redding, S. V. Carr, O. Sobolev, A. Schneidewind, E. Faulhaber, L. W. Harriger, S. Li, X. Lu, D.-X. Yao, T. Das, A. V. Balatsky, Th. Br\"uckel, J. W. Lynn, and P. Dai, Phys. Rev. B {\bf 88}, 064504 (2013).
	
	\bibitem{Hanaguri10} T. Hanaguri, S. Niitaka, K. Kuroki, H. Takagi, Science {\bf 328},  474 (2010).
	
	\bibitem{Coleman11} S. Sykora and P. Coleman, Phys. Rev. B {\bf 84}, 054501 (2011).

	\bibitem{Hirschfeld15} P. J. Hirschfeld, D. Altenfeld, I. Eremin, and I. I. Mazin, Phys. Rev. B {\bf 92}, 184513 (2015).
	
	\bibitem{Sawatzky} H. Wadati, I. Elfimov, and G. A. Sawatzky, Phys. Rev. Lett. \textbf{105}, 157004 (2010).
	
	\bibitem{Noat10} Y. Noat, T. Cren, V. Dubost, S. Lange, F. Debontridder, P. Toulemonde, J. Marcus, A. Sulpice, W. Sacks, and D. Roditchev, J. Phys. Cond. Matter \textbf{22}, 465701 (2010).
		
	\bibitem{Dai13} Y. M. Dai, B. Xu, B. Shen, H. H. Wen, X. G. Qiu, and R. P. S. M. Lobo, EPL \textbf{104}, 47006 (2013).
	
	\bibitem{Nakajima10} Y. Nakajima, T. Taen, Y. Tsuchiya, T. Tamegai, H. Kitamura, and T. Murakami, Phys. Rev. B {\bf 82}, 220504(R) (2010).
	
	\bibitem{Tarantini10} C. Tarantini, M. Putti, A. Gurevich, Y. Shen, R. K. Singh, J. M. Rowell, N. Newman, D. C. Larbalestier, P. Cheng, Y. Jia, and H. H. Wen, Phys. Rev. Lett. \textbf{104}, 087002 (2010).
	
	\bibitem{Beek13} C. J. van der Beek, S. Demirdis, D. Colson, F. Rullier-Albenque, Y. Fasano, T. Shibauchi, Y. Matsuda, S. Kasahara, P. Gierlowski, and M. Konczykowski, JoP Conference Series \textbf{449} (2013) 012023.
	
	\bibitem{Wang13} Y. Wang, A. Kreisel, P. J. Hirschfeld, and V. Mishra, Phys. Rev. B {\bf 87}, 094504 (2013).
	
	\bibitem{Efremov11}
	D. V. Efremov, M. M. Korshunov, O. V. Dolgov, A. A. Golubov, and P. J. Hirschfeld,  Phys. Rev. B \textbf{84}, 180512 (2011).
	
	\bibitem{Mizukami14} Y. Mizukami, M. Konczykowski, Y. Kawamoto, S. Kurata, S. Kasahara, K. Hashimoto, V. Mishra, A. Kreisel, Y. Wang, P. J. Hirschfeld, Y. Matsuda, and T. Shibauchi, Nat. Commun. \textbf{5}, 5657 (2014).
	
	\bibitem{Efremov13}
	D. V. Efremov, A. A. Golubov, and O. V. Dolgov,  New J. Phys. \textbf{15}, 013002 (2013).
	
	\bibitem{Golubov1995} A. A. Golubov and I. I. Mazin, Physica C 243, 153 (1995).
	
	\bibitem{Golubov1997} A. A. Golubov and I. I. Mazin, Phys. Rev. B \textbf{55}, 15146 (1997).
	
	\bibitem{Jin2003}
	B. B. Jin, T. Dahm, A. I. Gubin, E.-M. Choi, H. J. Kim, S.-I. Lee, W. N. Kang, and N. Klein, Phys. Rev. Lett. \textbf{91}, 127006 (2003).
	
	\bibitem{DresselGruener}
	M. Dressel and G. Gr\"uner, \textit{Electrodynamics of Solids} (Cambridge University Press, Cambridge, 2002).
	
	\bibitem{Aguilar10}
	R. V. Aguilar, L. S. Bilbro, S. Lee, C. W. Bark, J. Jiang, J. D. Weiss, E. E. Hellstrom, D. C. Larbalestier, C. B. Eom, and N. P. Armitage,  Phys. Rev. B \textbf{82}, 180514 (2010).
	
	\bibitem{Fischer10}
	T. Fischer, A. V. Pronin, J. Wosnitza, K. Iida, F. Kurth, S. Haindl, L. Schultz, B. Holzapfel, and E. Schachinger, Phys. Rev. B \textbf{82}, 224507 (2010).
	
	\bibitem{Nicol} E. J. Nicol and J. P. Carbotte, Phys. Rev. B \textbf{44}, 7741 (1991).
	
	\bibitem{Iida09}
	F. Kurth, E. Reich, J. H\"anisch, A. Ichinose, I. Tsukada, R. H\"uhne, S. Trommler, J. Engelmann, L. Schultz, B. Holzapfel, and K. Iida, Appl. Phys. Lett. \textbf{102}, 142601 (2013).
	
	\bibitem{Gorshunov05} B. P. Gorshunov, A. Volkov, I. E. Spektor, A. S. Prokhorov, A. A. Mukhin, M. Dressel, S. Uchida, and A. Loidl, Int. J. Infrared Millimeter Waves {\bf 26}, 1217 (2005).
	
	\bibitem{Pracht13}
	U. S. Pracht, E. Heintze, C. Clauss, D. Hafner, R. Bek, D. Werner, S. Gelhorn, M. Scheffler, M. Dressel, D. Sherman, B. Gorshunov, K. S. Il'in, D. Henrich, and M. Siegel, IEEE Trans. THz Sci. Technol. {\bf 3}, 269 (2013).
	
	\bibitem{SupplementalMaterials}
	Supplementary Materials.
	
	\bibitem{THz}
	S. Zapf, B. Gorshunov, D. Wu, E. Zhukova, V. S. Nozdrin, S. Haindl, K. Iida, and M. Dressel, J. Supercond. Nov. Magn. \textbf{21}, 1557 (2013).
	
	\bibitem{Popovich10} P. Popovich, A. V. Boris, O. V. Dolgov, A. A. Golubov, D. L. Sun, C. T. Lin, R. K. Kremer, and B. Keimer, Phys. Rev. Lett. \textbf{105}, 027003 (2010).
	
	\bibitem{Charnukha11} A. Charnukha, O. V. Dolgov, A. A. Golubov, Y. Matiks, D. L. Sun, C. T. Lin, B. Keimer, and A. V. Boris, Phys. Rev. B \textbf{84}, 174511 (2011).
	
	\bibitem{TcRRR}
	The residual resistivity ratio is calculated from the inverse optical conductivity at $\nu = \unit[5]{cm^{-1}}$: $RRR = \sigma_1(\unit[30]{K}) / \sigma_1(\unit[295]{K})$; the critical temperature was defined as the abrupt onset of frequency-dependent changes in the measured phase-shift, which is typical for entering the superconducting state. Since the transition has become too weak after the sixth irradiation to ultimately determine  $T_\text{c}$ from the optical data, after the last irradiation cycle the magnetic susceptibility was measured by V. Grinenkoa to add another point at higher irradiation doses.	
	
	
\end{thebibliography}
\end{document}